\documentclass{ws-rv9x6}
\usepackage{subfigure}   
\usepackage{ws-rv-thm}   
\usepackage{ws-rv-van}   
\usepackage{xspace}
\usepackage{hyperref}
\usepackage{amsmath}
\usepackage{upgreek}
\usepackage{mathrsfs}
\usepackage{breqn}

\makeindex
\newcommand{\pp}{\ensuremath{\rm pp}\xspace}
\newcommand{\st}{\ensuremath{S_{\rm T}}\xspace}
\newcommand{\so}{\ensuremath{S_{\rm 0}}\xspace}
\newcommand{\pt}{\ensuremath{p_{\rm{T}}}\xspace}
\newcommand{\nmpi}{\ensuremath{N_{\rm MPI}}\xspace}
\newcommand{\sppt}[1]{\ensuremath{\sqrt{s} = #1 \text{\,TeV}}\xspace}

\begin{document}

\chapter[Event Shapes at Hadron Colliders]{Experimental results on event shapes at hadron colliders}\label{ra_ch1}

\author[A. Ortiz]{Antonio Ortiz}

\address{Instituto de Ciencias Nucleares, Universidad Nacional Aut\'onoma de M\'exico,
Apartado Postal 70-543, M\'exico D.F. 04510, M\'exico \\
antonio.ortiz@nucleares.unam.mx}

\begin{abstract}

In this paper a review on event shapes at hadron colliders, mainly focused on experimental results, is presented. Measurements performed at the \textsc{Tevatron} and at the \textsc{Lhc}, for the soft and hard regimes of \textsc{Qcd}, are reviewed. The potential applications of event shapes for unveiling the origin of collective-like phenomena in small collision systems as well as for testing p\textsc{Qcd} predictions are discussed. 
\end{abstract}
\body


\section{Introduction}\label{ra_sec1}

In the hard regime of Quantum ChromoDynamics (\textsc{Qcd}), jets are produced from parton (quarks or gluons) scatterings at large transverse momentum (\pt). Since in that case the coupling constant ($\alpha_{\rm s}$) is small, precise perturbative \textsc{Qcd} (p\textsc{Qcd}) calculations are doable. In contrast, for processes involving low momentum transfer \textsc{Qcd}-inspired phenomenological models are built up.

Event shapes can be used to  test \textsc{Qcd} mainly because, by construction, they are collinear  and  infrared  safe  observables, i.e. they do not change their
value  when  an  extra  soft  gluon  is  added  or  if  a  parton  is  split  into  two  collinear  partons. This is a necessary condition for the cancellation of real and virtual divergences associated with such emissions, and therefore for making finite p\textsc{Qcd} predictions~\cite{Dasgupta:2003iq}.

Several studies using event shapes were performed in $e^{+}e^{-}$ annihilations, for example, in $e^{+}e^{-}$ at $\sqrt{s}=$22-44\, GeV~\cite{MovillaFernandez:1997fr}, p\textsc{Qcd} calculations (combining next-to-leading-order $\mathcal{O}(\alpha_{\rm s}^{2})$ and next-to-leading logarithms (NLLA) precisions) were combined with data for extracting the energy dependence of $\alpha_{\rm s}$. In the case of $\mathcal{O}(\alpha_{\rm s}^{2})$, the expression for a given event shape observable, $y$, generally has the form:
\begin{equation}
R (y)=1+A(y)\left( \frac{ \alpha_{\rm s} }{2\pi} \right)+B(y)\left( \frac{ \alpha_{\rm s} }{2\pi} \right)^{2},
\end{equation}
where $R(y)$ is the cumulative cross-section of $y$ normalised to the lowest order Born cross-section. While the analogous expression for NLLA calculations has the form:
\begin{equation}
R (y)=   \left(  1+C_{1}\left( \frac{ \alpha_{\rm s} }{2\pi} \right)+C_{2}\left( \frac{ \alpha_{\rm s} }{2\pi} \right)^{2} \right)\exp\left( Lg_{1}\left(  \frac{ \alpha_{\rm s} }{2\pi} L \right) + g_{2}\left(  \frac{ \alpha_{\rm s} }{2\pi} L \right) \right) 
\end{equation}
where $L=\ln(1/y)$. Clearly, from fitting the predictions to the data (corrected to the parton level) $\alpha_{\rm s}$ can be determined~\cite{BETHKE199854}.  These features allowed event shapes to be among the most studied  \textsc{Qcd} observables~\cite{Banfi:2001bz}, both theoretically and experimentally, not only in $e^{+}e^{-}$  but also in deep inelastic scattering.

For example, sphericity was used at \textsc{Slac} (Stanford Linear Accelerator Center) to testify to the existence of jets in $e^{+}e^{-}$ processes at center-of-mass energies up to 7.4\,GeV~\cite{PhysRevLett.35.1609}. Moreover, event shape variables were valuable tools which allowed the gluon discovery in processes where three prolonged jets were produced at energies up to 31.5\,GeV in the center-of-mass~\cite{PhysRevLett.43.830,BERGER1979418,BRANDELIK1979243}.  Since, event shape variables are defined with partons and they are measured in terms of visible particles, it is possible to use them for studying corrections due to hadronization effects~\cite{Kluth:2000km,ref1}.

In $e^{+}e^{-}\rightarrow {\rm q \bar{q}} \rightarrow {\rm hadrons}$ events, the particle production is studied with respect to the (unknown)  ${\rm q \bar{q}}$ axis, which can be approximated by the event shape axis. Clearly, the three components of the particle momentum are required in the calculation. In contrast, at hadron colliders since the particle production originated by the hadronic matter which interacts encodes the physics of interest, the event shape axis has to be searched in the plane perpendicular to the beam axis. The radiation perpendicular to the plane formed by the main hard scattering (event shape axis) and the beam axis would be sensitive to soft physics allowing to test \textsc{Qcd}-inspired phenomenological models.


Since at hadron colliders, p\textsc{Qcd} calculations are available for a vast number of event shapes~\cite{Banfi:2010xy,Becher:2015lmy}, they have been used to study different aspects of \textsc{Qcd}~\cite{Chatrchyan2013238,Khachatryan2014}. In addition, the shape of the events has also been proposed for discriminating events where black-holes (BH) are produced, or in supersymmetry (\textsc{Susy}) searches~\cite{Chatterjee:2012qt}. For example, BH events are characterized by higher sphericity than that expected in \textsc{Susy} processes~\cite{Roy:2008we}.

At hadron colliders, \textsc{Qcd} has been extensively tested in the perturbative regime, where precise calculations can be done. This is not the case of non-perturbative \textsc{Qcd}, therefore, the soft regime could represent an opportunity for the discovery of new phenomena. In this regard, the study of atypical events, like the ``hedgehogs''\footnote{High transverse energy (321\,GeV) isotropically distributed within a large acceptance ($\eta$$<$4).} observed in ${\rm p\bar{p}}$ collisions at \sppt{1.8} by the Collider Detector at \textsc{Fermilab} (CDF), was proposed  to unveil the nature of soft particle production at the  Large Hadron Collider (\textsc{Lhc}) energies~\cite{Quigg:2010ew}. Similar events were also reported at lower energies by the \textsc{Ua}1 Collaboration~\cite{Albajar1987}. Clearly,  ``hedgehog'' enriched samples could be easily achieved by running an event shape selection imposing a low \pt threshold.
 
Moreover, motivated by the discovery at the \textsc{Lhc} of collective-like phenomena in small collision systems (\pp and p-A collisions)~\cite{Loizides:2016tew,Bala:2016hlf},  the study of event shapes for soft physics has been encouraged. Since for the heavy-ion programme it is important to establish whether or not the strongly interacting Quark-Gluon-Plasma (s\textsc{Qgp}) is formed in small systems, where, the event shape analysis is a promising tool for controlling the jet bias in high multiplicity \pp events~\cite{Cuautle:2014yda,Cuautle:2015kra}.  

In this paper, a review on the measurements of event shapes at hadron colliders will be presented covering the soft and hard regimes. Data from different experiments will be used, going from \sppt{0.9} up to the largest energies reached by the Large Hadron Collider.

The paper is organised as follows, in Section~\ref{sec:2} the definitions of the different event shape variables are presented, the applications in the analyses of \pp data are discussed in Sections~\ref{sec:3} and \ref{sec:4} for the hard and soft regimes of  \textsc{Qcd}, respectively. An outlook is then displayed in Section~\ref{sec:5}.

\section{Definitions}
\label{sec:2}

As discussed in the introduction, at hadron colliders event shapes are defined in the transverse plane, i.e. in the plane perpendicular to the beam axis\footnote{It is worth mentioning that the \textsc{Atlas} Collaboration has also measured sphericity, but using jets and the full momentum tensor of the event, i.e., considering the component along the beam axis~\cite{Aad:2012np}. }. The objects (particles or jets) which participate in the calculation are restricted to some kinematic regions imposed by the detectors. For example, in the case of the transverse sphericity (\st)\cite{Abelev:2012sk} defined by the \textsc{Alice} Collaboration, only charged particles at mid-pseudorapidity ($|\eta|$$<$0.8) and with transverse momenta greater than 0.5\,GeV/$c$ are considered. For the transverse sphericity to be defined, the transverse momentum matrix, $\textbf{S}$, should be first diagonalized:

\begin{equation}
\textbf{S} =  \frac{1}{\sum_{i} p_{{\rm T},i}}  \sum_{i} \frac{1}{p_{{\rm T},i}}
\left( {\begin{array}{*{20}c}
   p_{{\rm x},i}^{2} &     p_{{\rm x},i}p_{{\rm y},i}  \\
  p_{{\rm y},i}p_{{\rm x},i}  &   p_{{\rm y},i}^{2} \\
 \end{array} } \right)
\end{equation}
where, $p_{{\rm T},i}$ is the transverse momentum of the $i$-th particle, being $p_{{\rm x},i}$ and $p_{{\rm y},i}$ the components along the $x$ and $y$ axes, respectively. The transverse sphericity is then defined in terms of the eigenvalues, $\lambda_{1}$$>$$\lambda_{2}$, as follows:

\begin{equation}
S_{\rm T} \equiv \frac{2\lambda_{2}}{\lambda_{1}+\lambda_{2}}
\end{equation}

Using the ratio of the smaller and larger eigenvalues, another event shape variable named $\mathcal{F}$-parameter can be defined as:
\begin{equation}
\mathcal{F} \equiv \frac{\lambda_{2}}{\lambda_{1}}
\end{equation}

On the other hand, transverse spherocity, originally proposed here~\cite{Banfi:2010xy} and recently studied in~\cite{Cuautle:2015kra} is defined for a unit vector $\boldsymbol{ \hat{\rm n}_{\rm s} }$ which minimizes the ratio:

\begin{equation}
S_{\rm 0} \equiv \frac{\pi^{2}}{4}  \underset{\bf \hat{n}_{\rm \bf{s}}}{\text{min}}  \left( \frac{\sum_{i}|{\vec p}_{{\rm T},i} \times { \bf \hat{n}_{\rm \bf{s}} }|}{\sum_{i}p_{{\rm T},i}}  \right)^{2}
\end{equation}

Concerning the thrust-related variables, the transverse thrust $\tau_{\rm T}$, defined as:

\begin{equation}
\tau_{\rm T} \equiv 1- \underset{\boldsymbol{ \hat{\rm n}_{\tau}}}{\text{max}} \frac{\sum_{i}|\vec{p}_{{\rm T},i}\cdot \boldsymbol{ \hat{\rm n}_{\tau}}|}{\sum_{i}p_{{\rm T},i}}
\end{equation}
considers the unit vector $\boldsymbol{ \hat{\rm n}_{\tau}}$ (thrust axis) that maximizes the sum, and thereby minimizes $\tau_{\rm T}$. The transverse thrust is more sensitive to the modelling of two- and three-jet topologies, while it is less sensitive to  \textsc{Qcd} modelling of larger jet multiplicities~\cite{Chatrchyan2013238}.

Using the direction given by $\boldsymbol{ \hat{\rm n}_{\tau}}$, the transverse region can be separated into an upper side $\mathcal{C}_{\rm U}$ (lower side $\mathcal{C}_{\rm L}$) consisting of all particles which satisfy $\vec{p}_{\rm T}$$\cdot$$\boldsymbol{ \hat{\rm n}_{\tau}}$$>$0 ($\vec{p}_{\rm T}$$\cdot$$\boldsymbol{ \hat{\rm n}_{\tau}}$$<$0). Moreover, the thrust axis $\boldsymbol{ \hat{\rm n}_{\tau} }$ and the beam direction $\boldsymbol{ \hat{z} }$ together define the event plane in which the primary hard scattering occurs.  Using this new axis, $\boldsymbol{ \hat{\rm n}_{\rm m} = \hat{\rm n}_{\tau} \times \hat{z}}$, the transverse thrust minor is defined as:

\begin{equation}
T_{\rm min} \equiv \frac{\sum_{i}|\vec{p}_{{\rm T},i}\cdot \boldsymbol{ \hat{\rm n}_{\rm m}}|}{\sum_{i}p_{{\rm T},i}}
\end{equation}
The observable $T_{\rm min}$ is a measure of the out-of-plane transverse momentum  and varies from zero, for an event entirely in the event plane, to $2/\pi$ for a cylindrically symmetric event.~\cite{Aaltonen:2011et}. It is interesting to see the interpretation of \so is similar to that for $T_{\rm m}$.

There are some common features among the event shapes defined before.
\begin{itemize}
\item The event shape is linear in particle momenta, therefore is a collinear safe quantity in p\textsc{Qcd}.
\item The lowest limit of the variable corresponds to the ``pencil-like'' structure (jetty-like structure for high multiplicity events).
\item The highest limit of the variable corresponds to the isotropic structure.
\end{itemize}

 \begin{figure}[t!]
\begin{center}
  \includegraphics[height=.30\textheight]{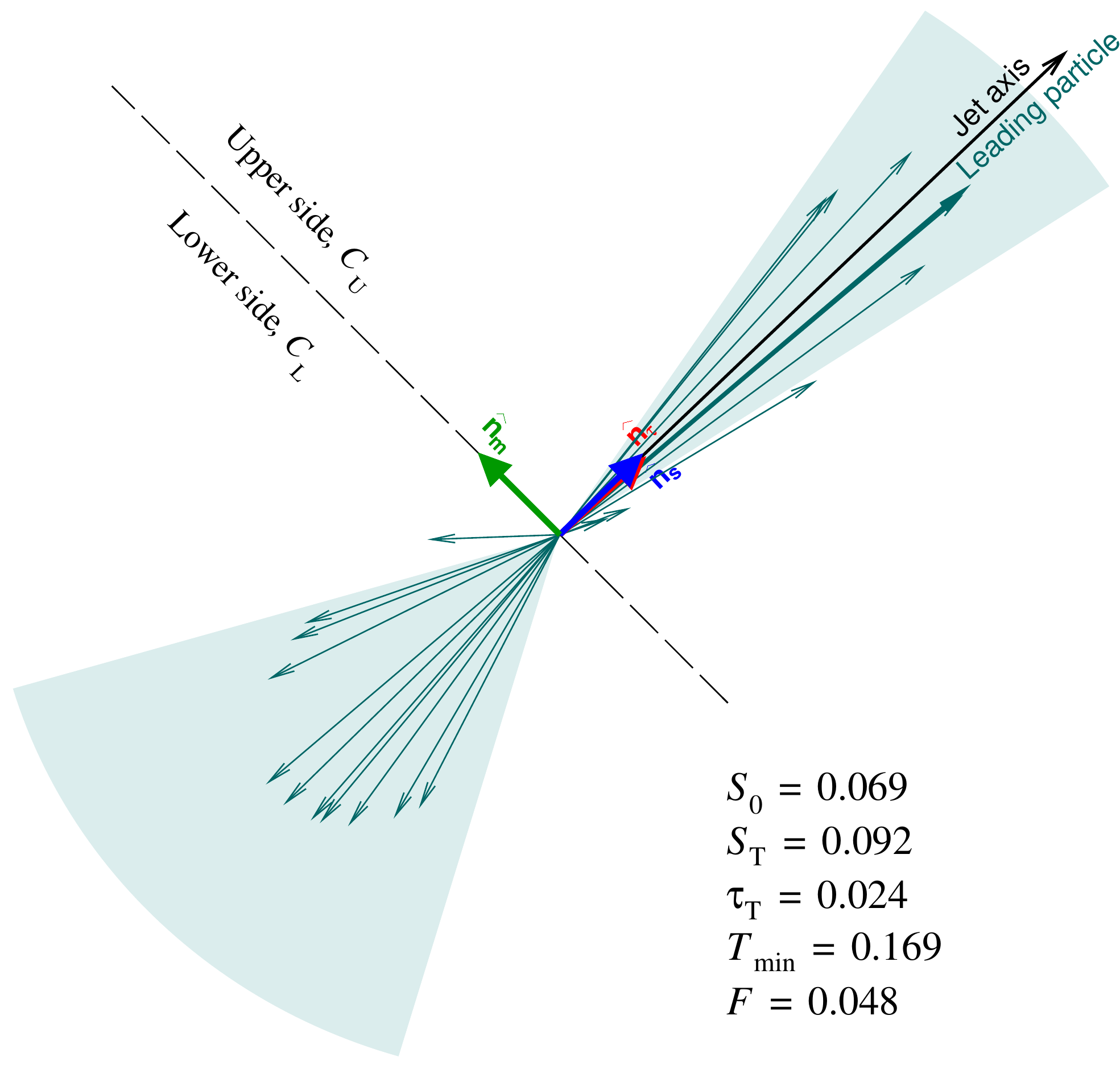}
  \includegraphics[height=.30\textheight]{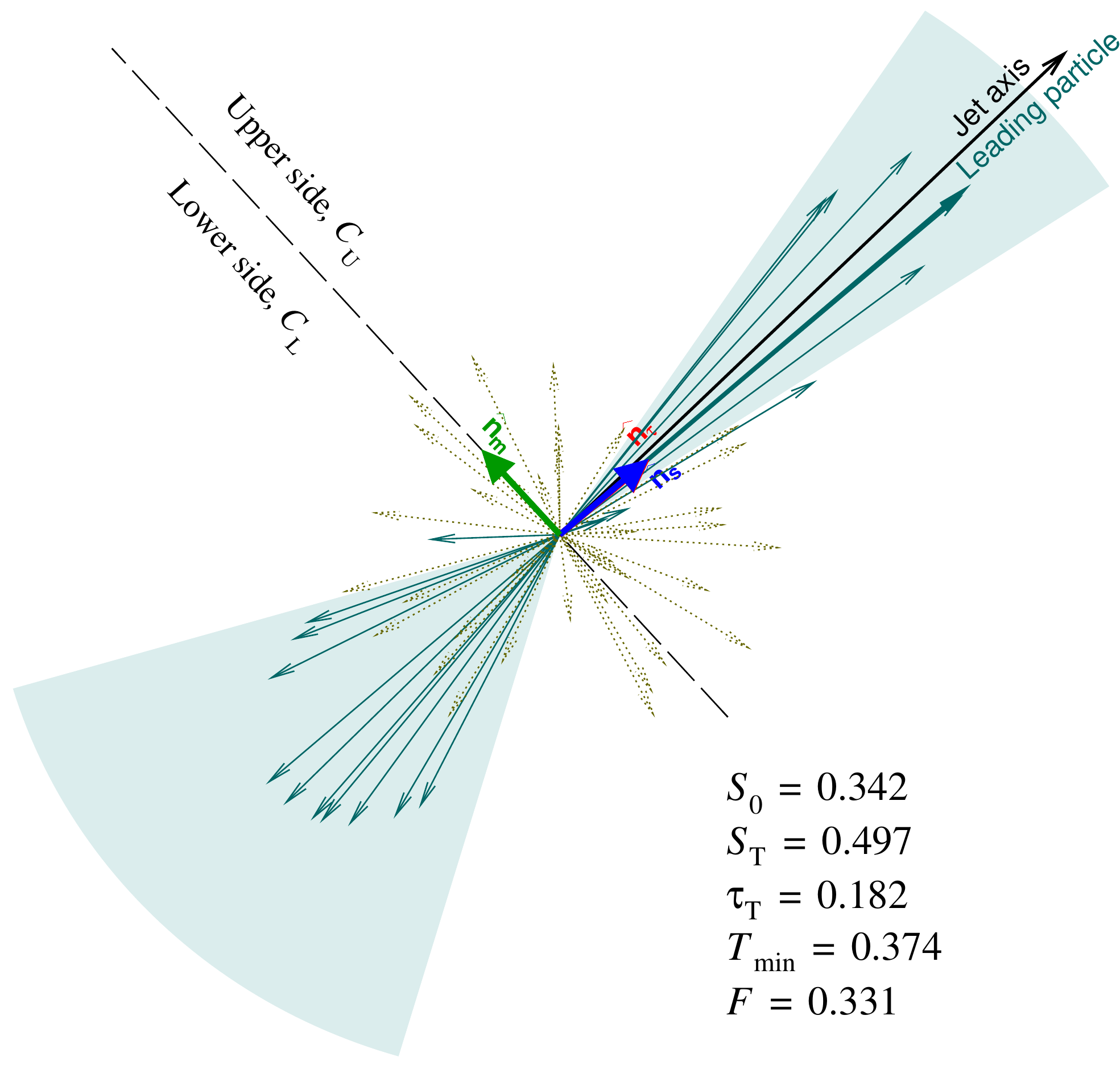}  
  \caption{(Color online) Sketch of a back-to-back jet without (left) and with (right) the underlying event, the information is projected in the plane perpendicular to the beam axis. The solid arrows indicate the particles within the jet, while, the dotted ones indicate particles belong to the underlying event. All the relevant event shape axes are also illustrated: spherocity (blue arrow), thrust (red arrow) and thrust minor (green arrow). The values of different event shapes are also displayed.}
\label{fig:hard:0}
\end{center}
\end{figure}

To illustrate the features of the different event shapes, Fig.~\ref{fig:hard:0} displays sketches of events as seen from the plane perpendicular to the beam axis. The left hand side figure exhibits the case when only particles belonging to a back-to-back jet are considered in the calculations of different event shapes. The different axes which were already discussed are also displayed. As expected, the thrust and spherocity axes are almost parallel to the jet axis. In addition, transverse sphericity, spherocity, thrust and $\mathcal{F}$ take values which are very close to zero, whereas, the variable which gives the out-of-plane \pt amounts to 0.169. When the underlying event (\textsc{Ue}), i.e. the soft component which accompanies the hard scattering is taken into account in the calculation (right hand side figure) of the event shapes:  \st, $\tau_{\rm T}$ and $\mathcal{F}$ are strongly affected, in contrast to \so and $T_{\rm m}$.  Moreover, the thrust and spherocity axes are now more aligned to the leading particle. It is worth recalling that for transverse spherocity is harder to reach the isotropic limit than for sphericity~\cite{Banfi:2010xy}, therefore, the discrimination power between isotropic soft events and symmetric multi-jet events might be the highest for spherocity.

Finally, the total jet broadenig ($B_{\rm tot}$) is defined as follows:
\begin{equation}
B_{\rm tot} \equiv B_{\rm U} + B_{\rm L}
\end{equation}
where U (L) refers to the upper (lower) side. The jet broadening variable in each region is defined as:
\begin{equation}
B_{\rm X} \equiv \frac{ \sum\limits_{i\in \mathcal{C}_{\rm X}} p_{{\rm T},i}\sqrt{ (\eta_{i}-\eta_{\rm X})^{2}+(\phi_{i}-\phi_{\rm X})^{2} } }{2\sum\limits_{i \in \mathcal{C}_{\rm X}} p_{{\rm T},i}}
\end{equation}
where the $i$-th particle is within the $\mathcal{C}_{\rm X}$ ($\mathcal{C}_{\rm U}$ or $\mathcal{C}_{\rm L}$) side. The pseudorapidity and azimuthal angles of the axes for the upper and lower sides are defined as follows:
\begin{equation}
\eta_{\rm X} \equiv \frac{ \sum\limits_{i\in \mathcal{C}_{\rm X}} p_{{\rm T},i}\eta_{i} }{\sum\limits_{i \in \mathcal{C}_{\rm X}} p_{{\rm T},i}}
\end{equation}
\begin{equation}
\phi_{\rm X} \equiv \frac{ \sum\limits_{i\in \mathcal{C}_{\rm X}} p_{{\rm T},i}\phi_{i} }{\sum\limits_{i \in \mathcal{C}_{\rm X}} p_{{\rm T},i}}
\end{equation}

\section{Hard QCD sector}
\label{sec:3}

This section aims to show that event shapes are still useful tools to test the validity of  \textsc{Qcd} prediction, and at the same time, to provide confidence in the current MC models for the description of the Standard Model processes and their application for determining background in new physics searches at hadron colliders.

\begin{figure*}[t!]
\begin{center}
\includegraphics[keepaspectratio, width=0.95\columnwidth]{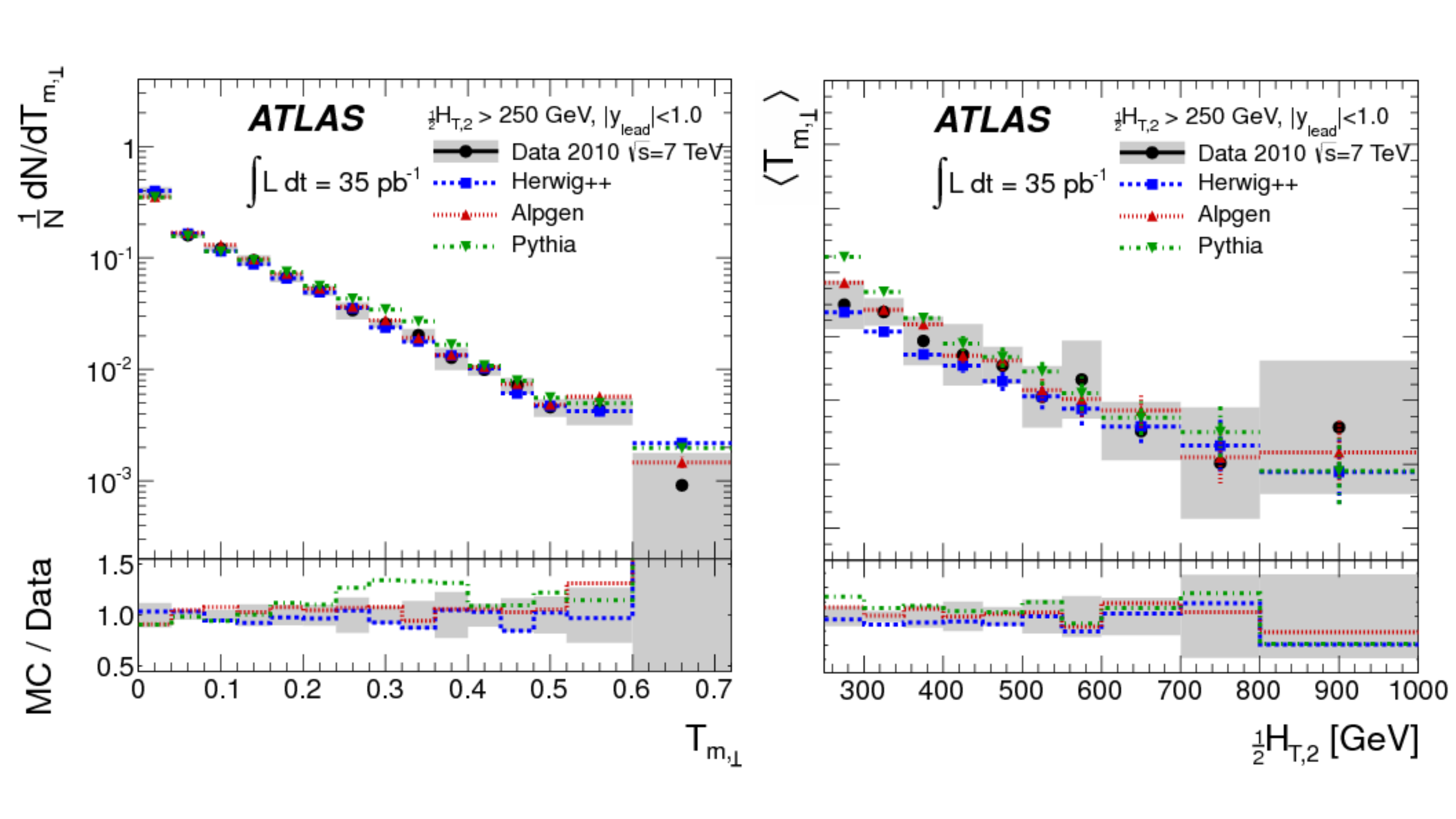}
\caption{\label{fig:hard:1} (Color online) Transverse thrust minor distribution for hard events ($\frac{1}{2}H_{\rm T,2}>250$\,GeV/$c$) (left) and average transverse thrust minor as a function of $\frac{1}{2}H_{\rm T,2}$ (right). Results for pp collisions at $\sqrt{s}=7$\,TeV are compared with different Monte Carlo generators. Figures reproduced from Ref.~\cite{Aad:2012np}.} 
\end{center}
\end{figure*}

At the Large Hadron Collider, the \textsc{Atlas} Collaboration has reported a study as a function of the scalar sum of the jet transverse momenta which is defined as:
\begin{equation}
\frac{1}{2}H_{\rm T,2} \equiv \frac{1}{2}(p_{\rm T,1}+p_{\rm T,2})
\end{equation}
where the subscript $i=1,2$ refers to the leading and sub-leading jet in the event~\cite{Aad:2012np}. Overall, the modelling of data by \textsc{Pythia}~6~\cite{Sjostrand:2006za} (tune \textsc{Perugia}~0~\cite{Skands:2010ak}) and \textsc{Alpgen}~\cite{Mangano:2002ea} are more accurate than that by \textsc{Herwing++}~\cite{Bahr:2008pv}. For example, Fig.~\ref{fig:hard:1} shows the transverse thrust minor distribution which within uncertainties is well described by the different models, albeit, the average thrust minor predicted by \textsc{Pythia} is slightly higher at low $\frac{1}{2}H_{\rm T,2}$ than in data. For the different event shapes, their mean values decrease with $\frac{1}{2}H_{\rm T,2}$ and the trend is well modeled by the Monte Carlo  (MC) generators. A similar study has been performed by the \textsc{Cms} Collaboration~\cite{Khachatryan:2011dx}. Although, the MCs used to compare the data are not the same to those used by \textsc{Atlas}~\cite{Aad:2012np}, the conclusions are indeed similar. 

\begin{figure*}[t!]
\begin{center}
\includegraphics[keepaspectratio, width=0.97\columnwidth]{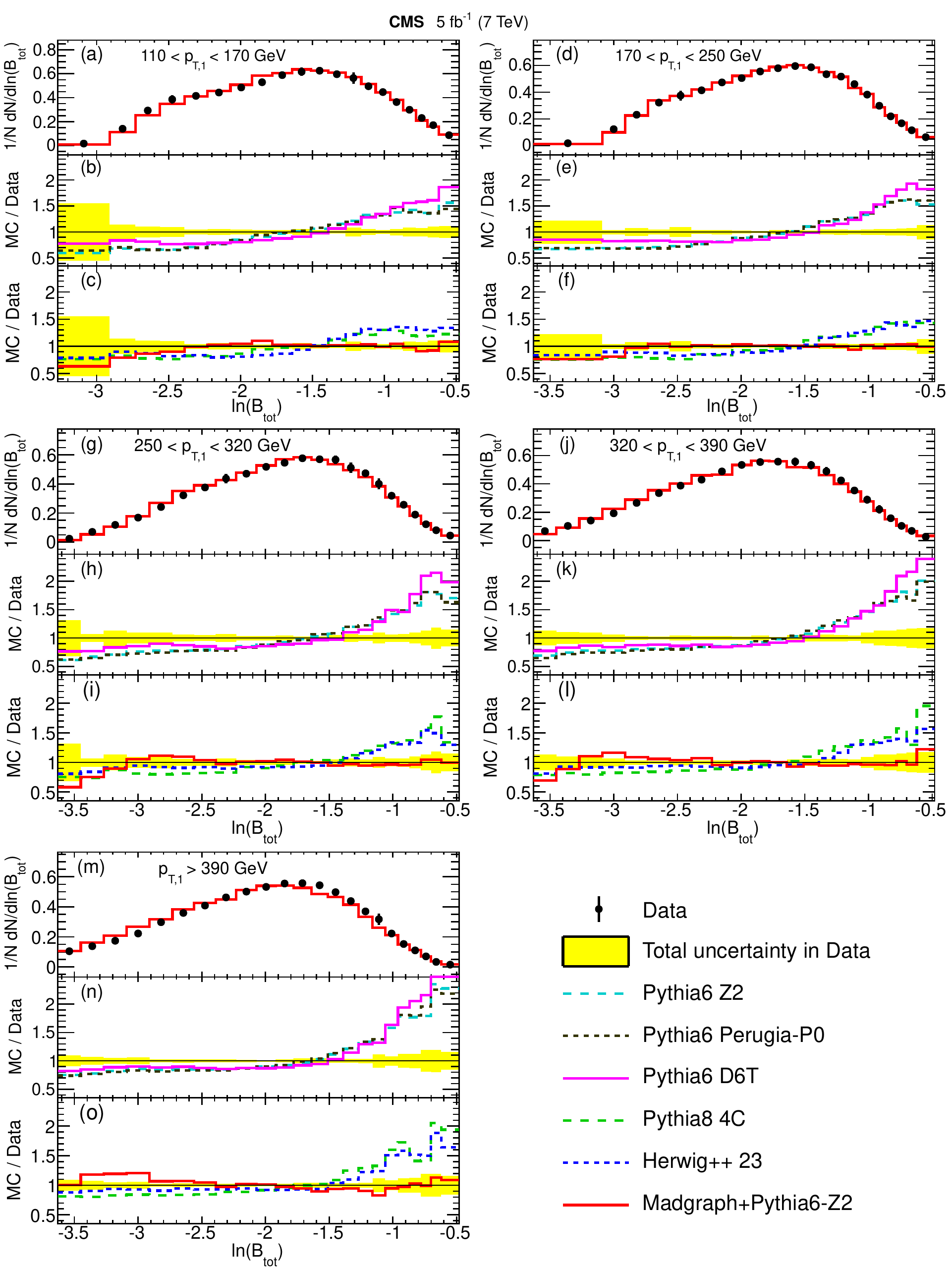}
\caption{\label{fig:hard:2} (Color online) Comparison between the jet broadening distributions in data and various Monte Carlo generators. Going from top left to bottom left, the leading jet \pt increases. Figures reproduced from Ref.~\cite{Khachatryan2014}.} 
\end{center}
\end{figure*}

The study was further extended using a larger dataset, in that case event shapes have been measured using jets as inputs in multi-jet events~\cite{Khachatryan2014}. It has been shown that event shapes that are more sensitive to the longitudinal energy flow show larger discrepancies between data and MC simulations. As an example, Fig.~\ref{fig:hard:2} shows the jet broadening distribution for various intervals of leading jet \pt. As discussed in~\cite{Khachatryan2014}, this observable is insensitive to the underlying event and hadronization details, however, a precise modeling of the matrix element calculations and parton showering is crucial in order to improve the description of the data. In this context,  within uncertainties the \textsc{MadGraph}~\cite{Alwall:2014hca} matrix-element (\textsc{Me}) calculator combined with \textsc{Pythia}~6 consistently reproduces all the distributions.

At the  \textsc{Tevatron}, the \textsc{Cdf} experiment has proposed a new quantity less dependent on the \textsc{Ue} aiming to have a more meaningful comparison between NLO+NLL parton-level predictions and the measured data~\cite{Aaltonen:2011et}. Separating the final state into hard and soft components and recognizing that the thrust axis is determined almost entirely by the hard component, the transverse thrust and transverse thrust minor can be written approximately as:

\begin{figure*}[t!]
\begin{center}
\includegraphics[keepaspectratio, width=0.95\columnwidth]{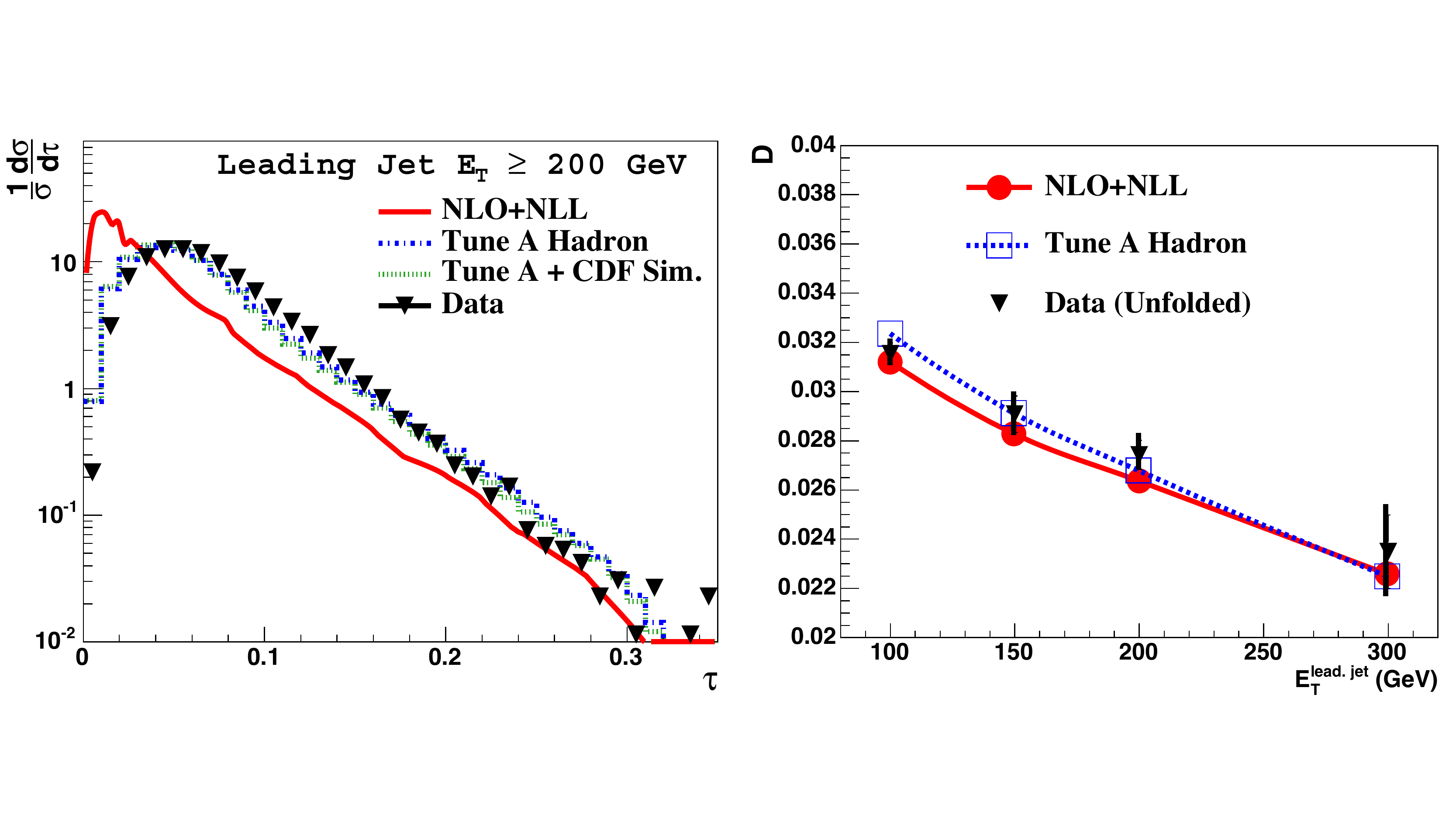}
\caption{\label{fig:hard:3} (Color online) The uncorrected distributions of transverse thrust for leading jet transverse energy greater than 200 GeV  (left) and the thrust differential as a function of the  transverse energy of the leading jet (right) measured in ${\rm p\bar{p}}$ collisions at $\sqrt{s}=1.96$\,TeV are compared with a parton level NLO+NLL calculations and with \textsc{Pythia}~6 at the hadron level. Figures reproduced from Ref.~\cite{Aaltonen:2011et}.} 
\end{center}
\end{figure*}

\begin{equation}
\tau_{\rm T} \approx  \frac{ \sum_{i}p_{{\rm T},i}^{\rm hard} - \underset{\boldsymbol{ \hat{\rm n}_{\tau}}}{\text{max}} \sum_{i}p_{{\rm T},i}^{\rm hard} | \cos{\phi^{\rm hard}_{i}} |}{\sum_{i}p_{{\rm T},i}^{\rm hard}+\sum_{j}p_{{\rm T},j}^{\rm soft}} + \frac{ \sum_{j}p_{{\rm T},j}^{\rm soft} (1-|\cos{\phi^{\rm soft}_{j}}|) }{\sum_{i}p_{{\rm T},i}^{\rm hard}+\sum_{j}p_{{\rm T},j}^{\rm soft}} 
\end{equation}

\begin{equation}
T_{\rm min} = \frac{\sum_{i}p_{{\rm T},i}^{\rm hard}|\sin{\phi_{i}^{\rm hard}}|}{\sum_{i}p_{{\rm T},i}^{\rm hard}+\sum_{j}p_{{\rm T},j}^{\rm soft}} + \frac{\sum_{j}p_{{\rm T},j}^{\rm soft}|\sin{\phi_{j}^{\rm soft}}|}{\sum_{i}p_{{\rm T},i}^{\rm hard}+\sum_{j}p_{{\rm T},j}^{\rm soft}}
\end{equation}
where $\phi^{\rm hard}$ ($\phi^{\rm soft}$) represents the azimuthal angle between the thrust axis and the hard (soft) component. Since for the soft underlying event:
\begin{equation}
 1-\tau_{\rm T}^{\rm soft}\approx T_{\rm min}^{\rm soft} \approx 2/\pi
 \end{equation}
the weighted difference between the mean values of the thrust and thrust minor: $\alpha \langle T_{\rm min} \rangle - \beta  \langle \tau_{\rm T} \rangle$, being $\alpha = 1-2/\pi$ and $\beta=2/\pi$, is a quantity less dependent on the \textsc{Ue}, the \textsc{Cdf} Collaboration defined the transverse thrust differential as follows:
\begin{equation}
D( \langle T_{\rm min} \rangle, \langle \tau_{\rm T} \rangle) \equiv \gamma_{\rm MC}(\alpha \langle T_{\rm min} \rangle - \beta  \langle \tau_{\rm T} \rangle)
\end{equation}
where the additional correction factor $\gamma_{\rm MC}$ is obtained from \textsc{Pythia}~6. Figure~\ref{fig:hard:3} shows the transverse thrust distribution measured in ${\rm p{\bar p}}$ collisions at \sppt{1.96} compared with NLO+NLL and \textsc{Pythia}~6 predictions, clearly NLO+NLL calculations deviate significantly from data and MC because they do not incorporate neither hadronization nor the underlying event. However,  \textsc{Pythia}~6 and the NLO+NLL calculations succeed in describing the data on the thrust differential.

Studies using  event shapes in Z+jets final states in \pp collisions at the \textsc{Lhc} have also been reported~\cite{Chatrchyan:2013tna,Aad:2016ria}. As discussed before, the comparisons of data with ``simple'' models like \textsc{Pythia}~6 have shown the importance of additional corrections from LO and NLO \textsc{Me} formulations. It has been shown that the level of agreement between \textsc{Pythia} and data improves for more highly boosted subset of events, e.g.,  where the transverse momentum of Z is greater than 150\,GeV/$c$. On the other hand, the MC models that combine multi-parton  \textsc{Qcd} LO \textsc{Me} interfaced to parton shower evolution tend to agree always with the data.

\section{Soft  QCD: Prospects for small systems}
\label{sec:4}

The main difference with respect to the studies using hard events, is that for soft physics the events shape calculation considers charged particles (tracks) as input. For example, the \textsc{Alice} Collaboration reported for the first time a study of transverse sphericity as a function of multiplicity considering tracks with transverse momentum above 0.5\,GeV/$c$ and within its Time Projection Chamber acceptance, $|\eta|$$<$0.8~\cite{Abelev:2012sk}.\footnote{Other studies using Monte Carlo (MC) generators have reduced the \pt cut in order to increase the sensitivity to the soft component.}

The study of soft physics in \pp and p-Pb collisions is attractive because in high multiplicity events unexpected new collective-like phenomena were recently discovered at the \textsc{Lhc}. In particular, for high-multiplicity proton--proton  and  proton--lead collisions, radial flow signals~\cite{Abelev:2013haa,Adam:2016dau}, long-range angular correlations~\cite{ABELEV:2013wsa,Khachatryan:2016txc}, and the strangeness enhancement~\cite{Adam:2016emw,Adam:2015vsf,Khachatryan:2016yru} 
have been reported. Those effects are well-known in heavy-ion collisions, where they are attributed to the existence of the s\textsc{Qgp}~\cite{Bala:2016hlf,Abelev:2014laa,Adam:2015kca}. 
Understanding the phenomena is crucial because, for heavy-ion physics, pp and p--Pb collisions have been used as the baseline (``vacuum'') to extract the genuine s\textsc{Qgp} effects. However, it is worth mentioning that no jet 
quenching effects have been found so far in p--Pb collisions~\cite{Adam:2015hoa}, suggesting that other mechanisms could also play a role in producing collective-like behavior in small collision systems~\cite{Ortiz:2015cma,Ortiz:2013yxa,Zakharov:2015gza}.  

\begin{figure*}[t!]
\begin{center}
\includegraphics[keepaspectratio, width=0.99\columnwidth]{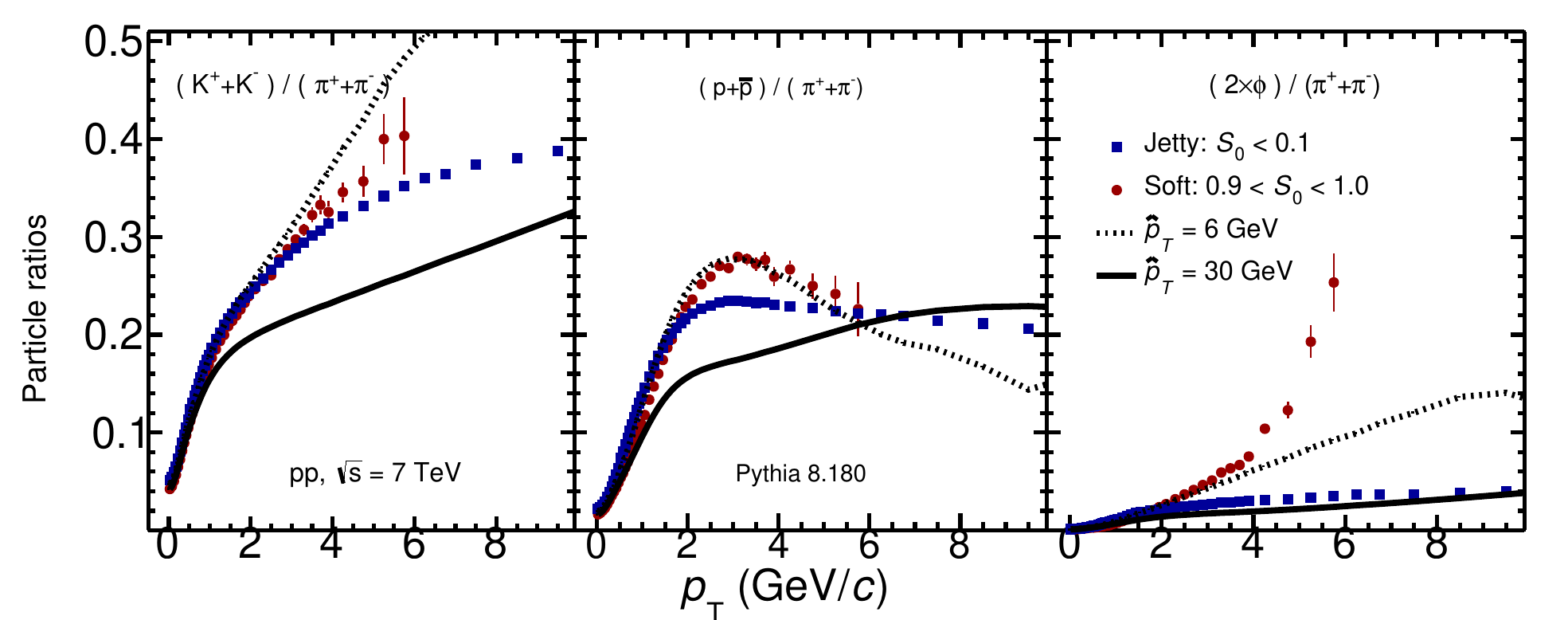}
\caption{\label{fig:soft:1} (Color online) Particle ratios as a function of \pt for jetty-like (squares) and soft (circles) events obtained in inclusive pp collisions simulated with \textsc{Pythia}~8. Results are compared to pp collisions where jets with transverse momentum ($\hat{p}_{\rm T}$) of 6 (dotted line) and 30 (solid line) GeV/$c$ are produced. Figure reproduced from Ref.~\cite{Cuautle:2014yda}.} 
\end{center}
\end{figure*}

The issue with the high multiplicity pp collisions is that a non-negligible part of the events may have a p\textsc{Qcd} origin, and, it has been demonstrated that jets may play an important role in the production of the collective-like effects observed in small systems~\cite{Ortiz:2016kpz,Cuautle:2015kra,Cuautle:2014yda}. Fortunately, the event shape analysis is a promising tool for controlling the jet bias in high multiplicity \pp events~\cite{Cuautle:2014yda,Cuautle:2015kra}.

A simulation study using inclusive \pp collisions (without any selection on multiplicity) has already unveiled a difference in the particle composition when one studies the jetty-like and isotropic events, separately. For example, Fig.~\ref{fig:soft:1} shows the \pt-integrated particle ratios (${\rm K}/\pi$, ${\rm p}/\pi$ and $\phi/\pi$) calculated for \pp collisions at \sppt{7} using \text{Pythia}~8~\cite{Sjostrand:2007gs}. For \pt below 2\,GeV/$c$ the ratios exhibit a depletion going from low to high \so, whereas, for larger \pt the ratios increase with spherocity. For \pt above 5-6 GeV/$c$ the proton-to-pion ratio in high \so events gets more similar to that for for low \so. The particle ratios which involve strange hadrons deviate each other, i.e. the values for jetty-like and isotropic events, for  $p_{\rm T}$$>$2\,GeV/$c$. These results are compared to those where the inelastic \pp interactions have a partonic $\hat{p}_{\rm T}$ of 6 or 30\,GeV/$c$, being this the most energetic p\textsc{Qcd} process within the event. The sample with 6\,GeV/$c$ jets is in a qualitatively good agreement with results for spherical events suggesting that this tool can be used to isolate the underlying event. On the other hand, the particle ratios decrease with increasing jet \pt.  It is worth noticing that a similar effect has been observed in the hadrochemistry measured in the so-called  ``bulk'' (outside the jet peak) and the jet regions in p--Pb and Pb--Pb collisions at the \textsc{Lhc}~\cite{Veldhoen:2012ge,XZhang:2014sva}. 

\begin{figure*}[t!]
\begin{center}
\includegraphics[keepaspectratio, width=0.99\columnwidth]{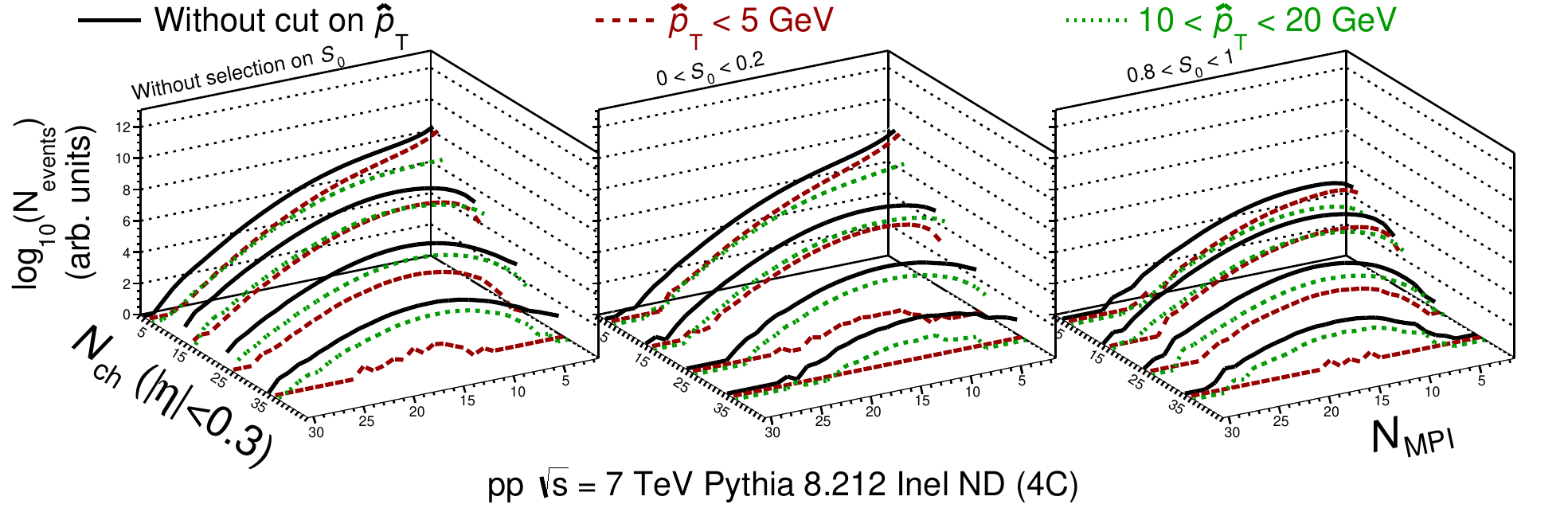}
\caption{\label{fig:soft:2} (Color online) Distributions of the number of multi-parton interactions as a function of the event multiplicity and leading partonic \pt ($\hat{p}_{\rm T}$) for non-diffractive (ND) inelastic \pp collisions at \sppt{7}. From left to right, inclusive, jetty-like events (low spherocity) and isotropic events (high spherocity) are shown. Figure reproduced form Ref.~\cite{Cuautle:2015fbx}.} 
\end{center}
\end{figure*}

Another potential application is related with the possibility of using the event shapes for selecting events with different number of multi-partonic interactions (MPI)~\cite{Cuautle:2015fbx,Cuautle:2015kra}. Figure~\ref{fig:soft:2} shows  the distribution of number of MPI (\nmpi) as a function of multiplicity for non-diffractive \pp collisions and for two extreme spherocity bins, $S_{\rm 0}$\,$<$\,$0.2$ (jetty-like) and $S_{\rm 0}$\,$>$\,$0.8$ (isotropic).  Also shown are the results for two intervals of the leading parton transverse momentum ($\hat{p}_{\rm T}$) of the event: $\hat{p}_{\rm T}$\,$<$\,$5$\,GeV and $10$\,$<$\,$\hat{p}_{\rm T}$\,$<$\,$20$\,GeV. We observe that the average number of multi-parton interactions increases with increasing multiplicity, slightly more for isotropic events. At low multiplicity the dominant events are those of low \nmpi and with $\hat{p}_{\rm T}$\,$<$\,$5$\,GeV/$c$. On the other hand, in high multiplicity events the number of events with $10$\,$<$\,$\hat{p}_{\rm T}$\,$<$\,$20$\,GeV is larger than that for $\hat{p}_{\rm T}$\,$<$\,$5$\,GeV. In \textsc{Pythia}, high multiplicity events are therefore produced via hard partonic scatterings~\cite{Abelev:2012sk}.  It is worth noticing that the event is isotropic or jetty-like within the restricted $\eta$-range which is considered for the calculation of the spherocity. This suggests that when selecting isotropic events we study enriched \textsc{Ue} samples, while jetty-like events are those which have poor underlying event activity within the acceptance under consideration.

\begin{figure*}[t!]
\begin{center}
\includegraphics[keepaspectratio, width=0.99\columnwidth]{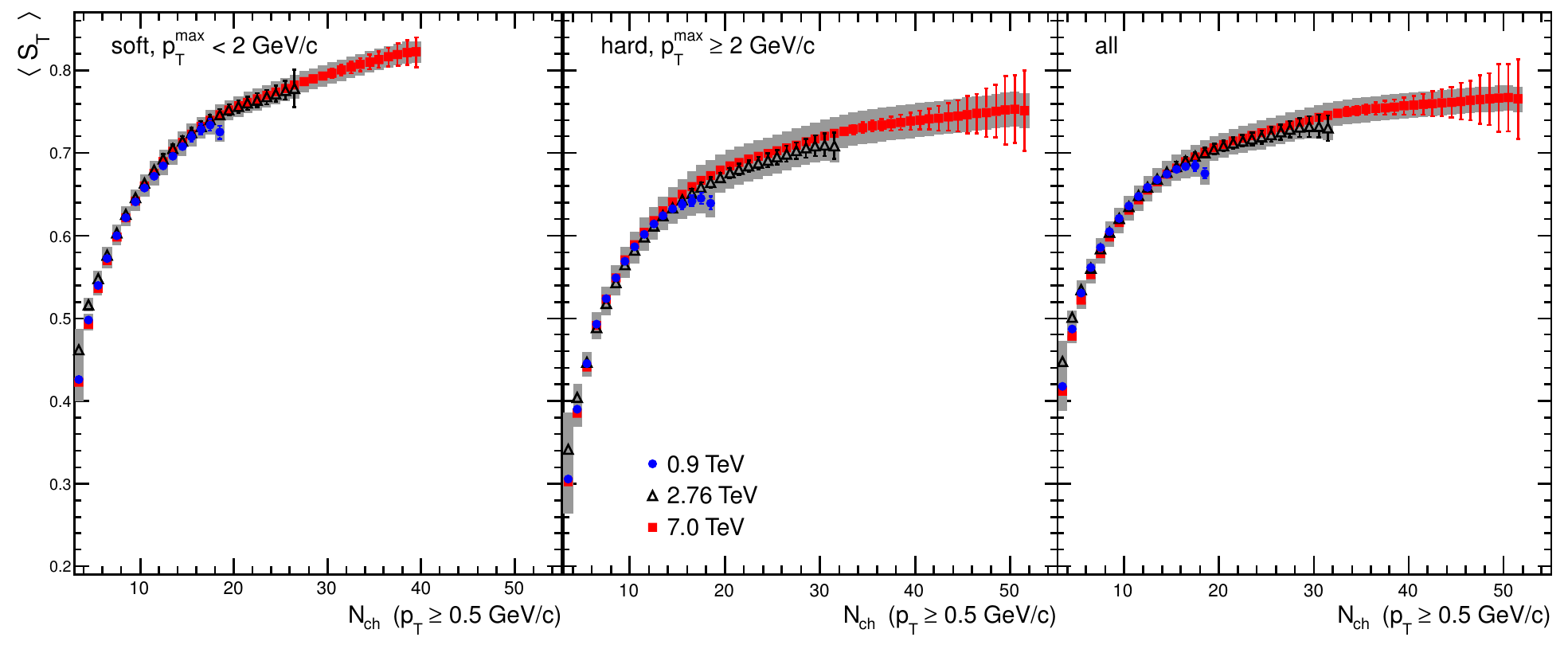}
\caption{\label{fig:soft:3} (Color online) Mean transverse sphericity versus multiplicity for inclusive (right),  ``hard'' (middle) and ``soft'' (left) \pp collisions at \sppt{0.9}, 2.76 and 7\,TeV. The statistical errors are displayed as error bars and the systematic uncertainties as the shaded area. Figure reproduced from Ref.~\cite{Abelev:2012sk}.} 
\end{center}
\end{figure*}

\begin{figure*}[t!]
\begin{center}
\includegraphics[keepaspectratio, width=0.99\columnwidth]{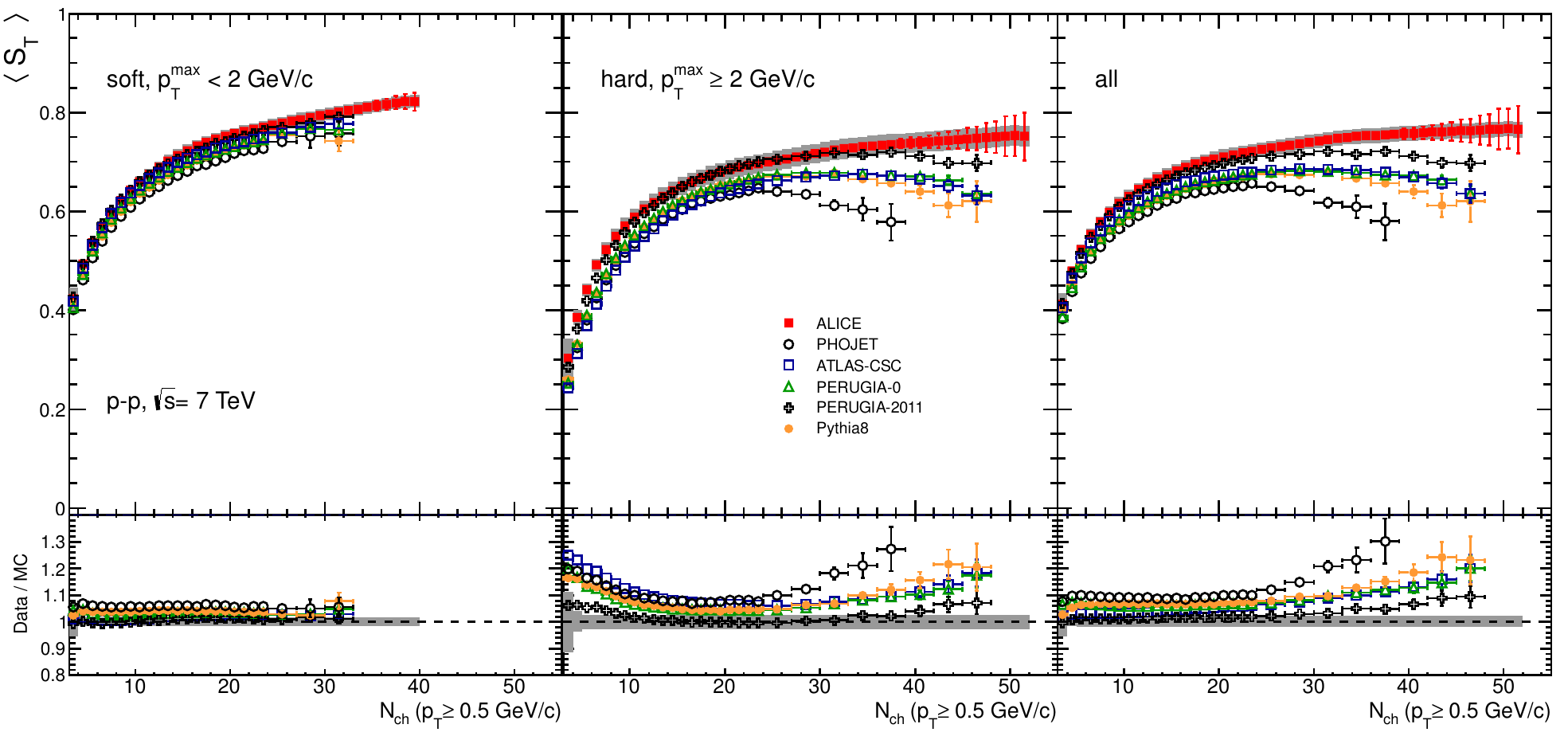}
\caption{\label{fig:soft:4} (Color online) Mean transverse sphericity versus multiplicity for inclusive (right),  ``hard'' (middle) and ``soft'' (left) \pp collisions at \sppt{7}. The \textsc{Alice} data are compared with different MC models: \textsc{Phojet}, \textsc{Pythia}~6 (tunes: \textsc{Atlas-Ccs}, \textsc{Perugia-0} and \textsc{Perugia-2011}) and \textsc{Pythia}~8. The statistical errors are displayed as error bars and the systematic uncertainties as the shaded area. Figure reproduced from Ref.~\cite{Abelev:2012sk}.} 
\end{center}
\end{figure*}

The first measurement at the \textsc{Lhc} on event shapes using MB events came from the \textsc{Alice} collaboration, which reported the average transverse sphericity as a function of the event multiplicity. Both quantities were computed at mid-pseudorapidity ($|\eta|$$<$0.8) using \pp data at $\sqrt{s}=0.9$, 2.76 and 7\,TeV~\cite{Abelev:2012sk}. The mean sphericity as a function of multiplicity was studied for different event classes selected using a cut on the maximum \pt of the event (leading particle), being the ``soft'' (``hard'') events those without (with) the leading particle having \pt above 2\,GeV/$c$. In Fig.~\ref{fig:soft:3}, the results are presented for the aforementioned event classes along with the inclusive case (``all'').  Like in other observables recently reported in the context of collectivity in small systems~\cite{Chatrchyan:2012qb,Adam:2016dau}, the average sphericity does not show dependence on $\sqrt{s}$. Instead, the physics seems to be encoded in the multiplicity. For ``soft'' events the rise is significantly steeper than in ``hard'' events, however, no indication of a reduction of the average transverse sphericity is drawn at high multiplicity for inclusive and ``hard'' \pp collisions. The comparison of these results with MC predictions (\textsc{Pythia}~6, \textsc{Pythia}~8 and \textsc{Phojet}~\cite{Engel:1995yda}) is quite illuminating, because as shown in Fig.~\ref{fig:soft:4} the MC average sphericity exhibits a reduction at high multiplicity, in inclusive and ``hard'' \pp collisions, whereas the sphericity in data stays constant or shows a little increase with multiplicity. In  \textsc{Qcd}-inspired MC generators, the prime mechanism to produce high multiplicity \pp collisions is related with the partonic interactions with large momentum transfer, therefore, the reduction may indicate an increase on the production of back-to-back jets. The deviation of the data from  \textsc{Qcd}-inspired predictions can be explained in the framework of the string percolation model~\cite{Bautista:2012tn} where the area covered by the strings increases with the size (multiplicity) and energy of the system leaving less room for hard scatterings~\cite{Andres:2012ma}. Elliptic flow has been also proposed to explain the differences observed at high multiplicity~\cite{Yan:2014vta}. A complementary study using jets was further performed by the \textsc{Cms} Collaboration, supporting the findings reported by \textsc{Alice}. Namely, the deviations from \textsc{Pythia} prediction at high multiplicity could be due to an apparent
reduction and softening of the jet yields~\cite{Chatrchyan:2013ala}.


\begin{figure}[t!]
\begin{center}
  \includegraphics[height=.231\textheight]{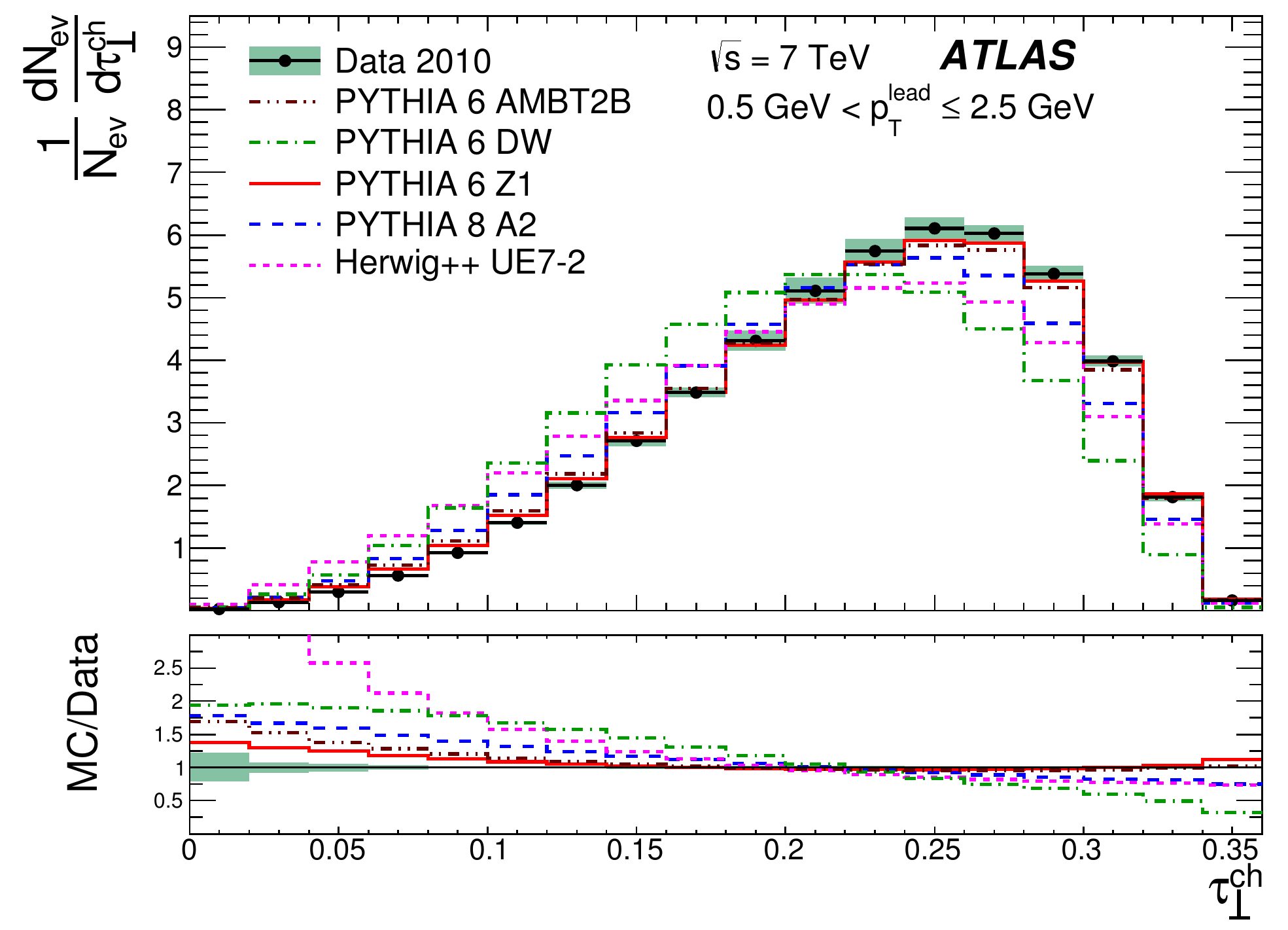}
  \includegraphics[height=.231\textheight]{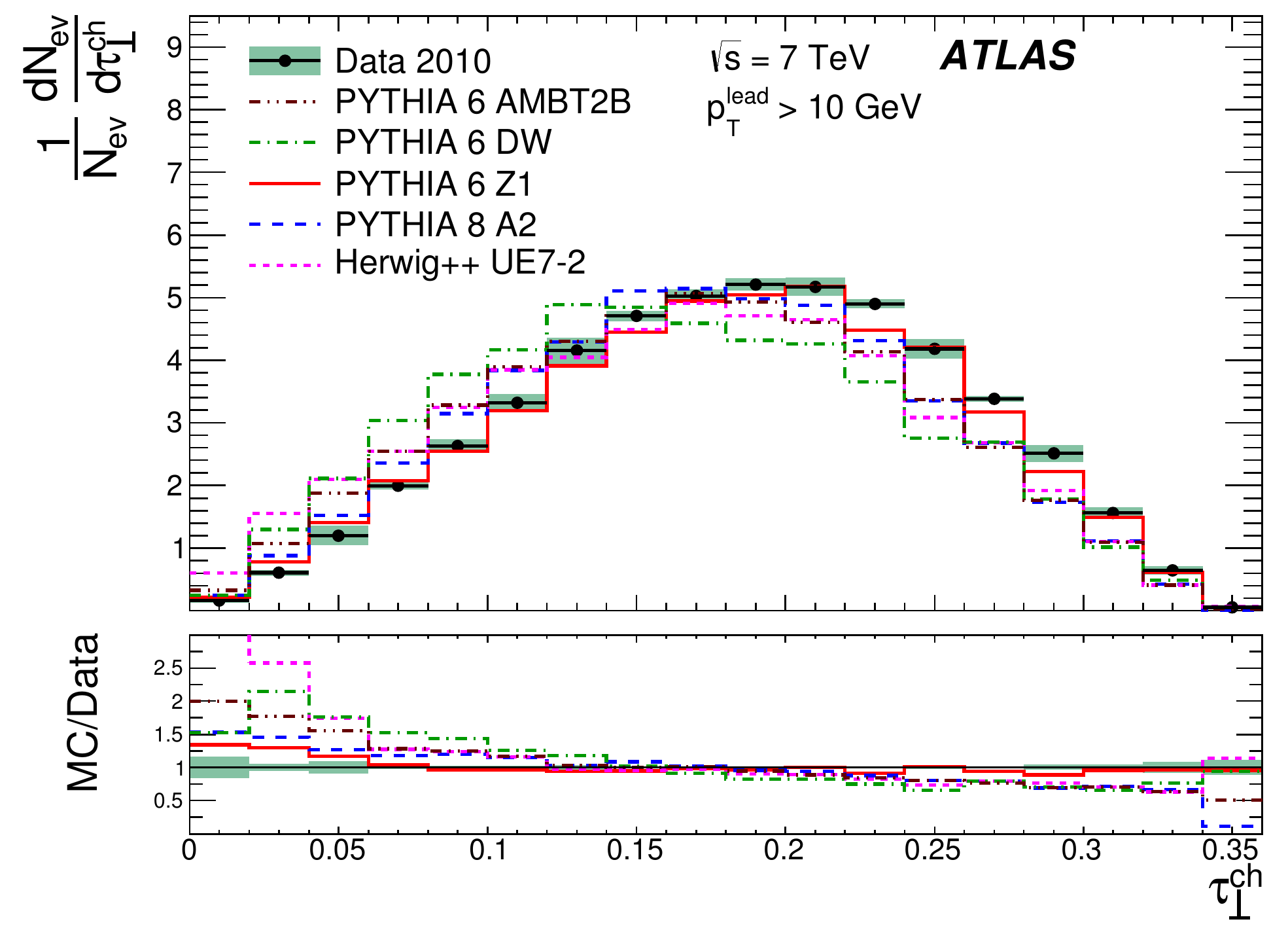}  
  \caption{(Color online) Transverse thrust distributions for \pp collisions at \sppt{7}. The results are presented for ``soft'' (left) and ``hard'' (right) events selected using a cut on the \pt of the leading particle. Data are compared with different MC  models: \textsc{Pythia}~6, \textsc{Pythia}~8 and \textsc{Herwing}++. Figures reproduced from Ref.~\cite{Aad:2012fza}.}
\label{fig:soft:5}
\end{center}
\end{figure}

The \textsc{Atlas} Collaboration reported a similar study but using transverse thrust, thrust minor and transverse sphericity, each defined for events having at least six charged particles ($|\eta|$$<$2.5 and $p_{\rm T}$$>$0.5\,GeV/$c$) for the calculation of the event shape~\cite{Aad:2012fza}. The distributions shown in Fig.~\ref{fig:soft:5} indicate a prevalence of spherical events in the  lower leading \pt intervals, and then a slight shift toward less spherical events and a broadening of the distributions is observed for events having larger leading particle \pt ($>$7.5\,GeV/$c$). The different event generators (\textsc{Pythia}~6~\cite{Sjostrand:2006za}, \textsc{Pythia}~8~\cite{Sjostrand:2007gs} and \textsc{Herwing}++~\cite{Bahr:2008pv}) which are compared with data share a similar feature, they overestimate (underestimate) the production of low (high) spherical events. Overall, the \textsc{Pythia}~6 tune Z1, tuned to the underlying event distributions at the \textsc{Lhc}, agrees the best with most of the distributions.  The \textsc{Pythia}~6 DW tune predictions
are consistently furthest from the data.  This is not unexpected as DW is tuned to reproduce the \textsc{Tevatron} data and does not agree with the measured charged particle multiplicity and \pt distributions at the \textsc{Lhc}.  For the different event shapes studied by \textsc{Atlas}, the evolution toward a more spherical event shape with increasing multiplicity is rapid initially and slows at higher multiplicities.

\begin{figure*}[t!]
\begin{center}
\includegraphics[keepaspectratio, width=0.55\columnwidth]{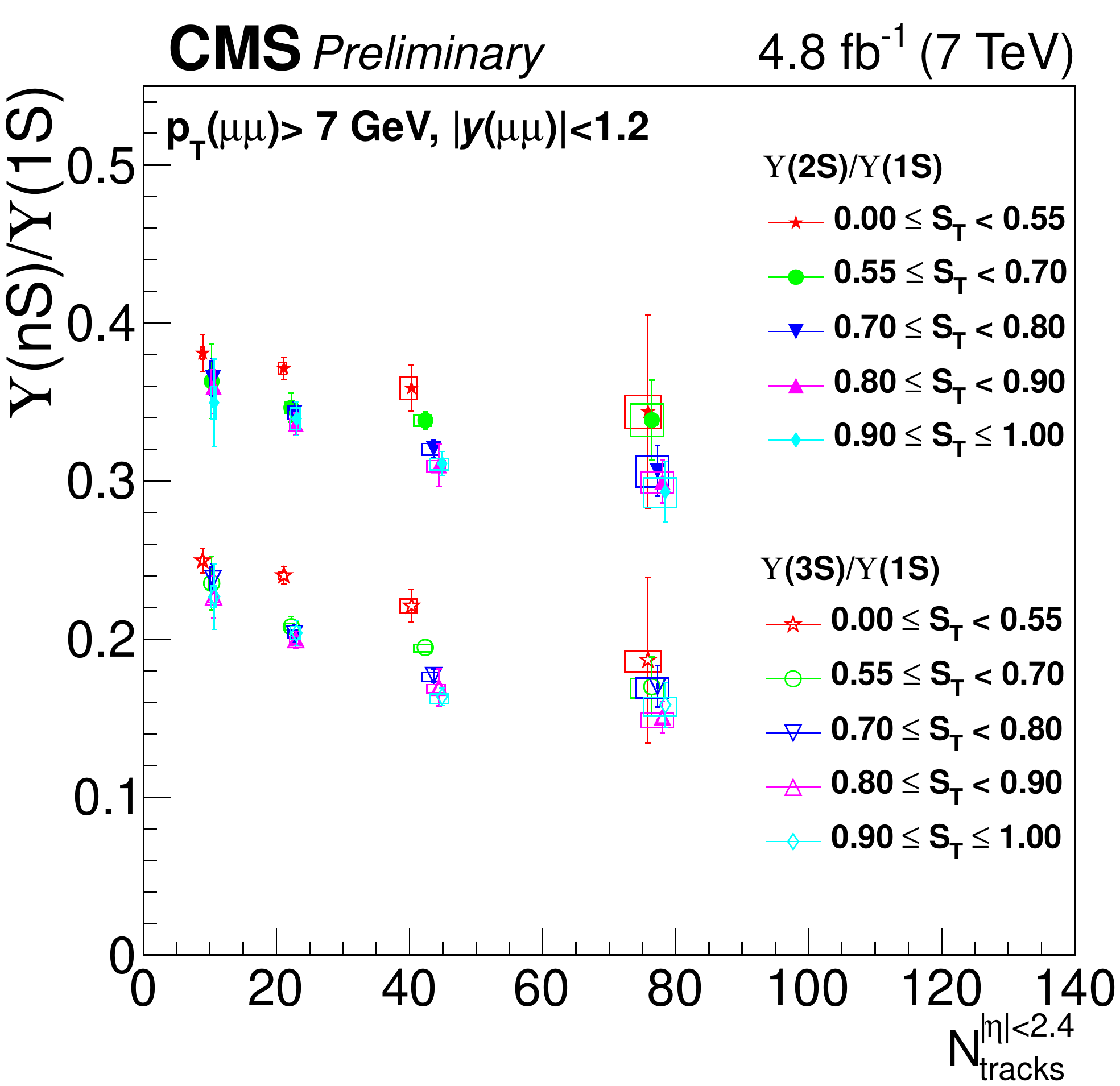}
\caption{\label{fig:soft:6} (Color online)  $\Upsilon(2{\rm S})/\Upsilon(1{\rm S})$ and $\Upsilon(3{\rm S})/\Upsilon(1{\rm S})$ as a function of the event multiplicity ($|\eta|$$<$2.4 and \pt$>$0.4\,GeV/$c$) in different intervals of sphericity. The $\Upsilon$ states satisfy \pt$>$7\,GeV/$c$ and $|y|$$<$1.2. Figure reproduced from Ref.~\cite{CMS-PAS-BPH-14-009}.} 
\end{center}
\end{figure*}

The event shape analysis at hadron colliders has been shown to be feasible for soft physics. Now the next step is to study the events cutting on the shape. In this way, at high multiplicities jets and the underlying event (or maybe s\textsc{Qgp} enhanced samples) can be analysed separately. The first attempts have been reported by the \textsc{Cms} Collaboration who has measured observables traditionally used to probe the s\textsc{Qgp} formation in heavy-ion collisions. Using \pp data at \sppt{7}, the ratios: $\Upsilon(2{\rm S})/\Upsilon(1{\rm S})$ and $\Upsilon(3{\rm S})/\Upsilon(1{\rm S})$ have been measured as a function of the charged particle multiplicity and for different sphericity classes~\cite{CMS-PAS-BPH-14-009}. The ratios exhibit a reduction with increasing multiplicity, being the $\Upsilon(2{\rm S})/\Upsilon(1{\rm S})$ ratios significantly higher than those for $\Upsilon(3{\rm S})/\Upsilon(1{\rm S})$. Moreover, in Fig.~\ref{fig:soft:6} we can see their sphericity dependences which show that the ratios are systematically lower (effect of $\approx$40\%) in spherical events than in jetty-like events. These observations resemble features of the sequential suppression of the $\Upsilon({\rm nS})$ states observed in heavy-ion collisions~\cite{Chatrchyan:2012lxa}. 

So far, the results are encouraging and therefore, further studies are desirable.

\section{Outlook}
\label{sec:5}

The event shape variables have been widely studied using hard events of ${\rm p\bar{p}}$ and pp collisions at the \textsc{Tevatron} and at the \textsc{Lhc}, respectively. The vast amount of data has allowed to the test the p\textsc{Qcd} predictions. Overall, a good agreement between data and theory is observed, the deviations are well understood as due to the lack of precise modelling of the matrix element calculations and parton showers. Moreover, the measured event shapes which were constructed to be less sensitive to the underlying event agree much better with the theoretical expectations.

For soft physics, the recent discovery of s\textsc{Qgp}-like signatures in small systems calls for new analysis techniques in order to control the p\textsc{Qcd} processes in high multiplicity events. As shown in this review, an analysis combining event shapes and multiplicity is a potential new direction in the study of the origin of the most intriguing phenomena seen at the \textsc{Lhc}. Two examples were presented. The event shape distributions at high multiplicity which showed that models underestimate the amount of isotropic events and overestimate the production of jetty-like events. Suggesting that in nature soft physics may dominate at high multiplicities. On the other hand, the preliminary study using $\Upsilon({\rm nS})$ states shows different $\Upsilon(2{\rm S})/\Upsilon(1{\rm S})$ and  $\Upsilon(3{\rm S})/\Upsilon(1{\rm S})$ ratios for jetty-like and isotropic events. Further studies are needed to established whether or not this is connected with the sQGP formation. Finally, the Monte Carlo results presented here encourage the development of analogous analyses using data.

\section*{Acknowledgements}

The author acknowledges Guy Pai\'c and Peter Christiansen for the critical reading of the manuscript and the valuable discussion and suggestions. Support for this work has been received by CONACyT under the grant number 280362.


\bibliographystyle{ws-ijmpe}
\bibliography{biblio_final}



\end{document}